# Coupled $\Lambda N - \Sigma N$ and $\Lambda NN - \Sigma NN$ systems and hyperon-nucleon interactions

K. Miyagawa[1], H.Kamada[2], W. Glöckle[2], H. Yamamura[1], T. Mart[3], C. Bennhold[4],

[1] Department of Applied Physics, Faculty of Science, Okayama University of Science, Okayama 700, Japan
[2] Institut für theoretische Physik II, Ruhr-Universität Bochum, D-44780 Bochum, Germany
[3] Jurusan, Fisika, FMIPA, Universitas Indonesia, Depok 16424, Indonesia
[4] Center for Nuclear Studies, Department of Physics, The George Washington University, Washington, D.C. 20052, USA

**Abstract.** This paper summarizes our studies of the hypertriton and describes our findings on the hyperon-nucleon interaction, especially on the role of the $\Lambda N - \Sigma N$ coupling and the strengths of the $S$-wave $YN$ interactions. We briefly comment on our ongoing analyses with electromagnetic probes.

## 1 Introduction

One of the most prominent features of the $\Lambda N$ interaction is the lack of the one-pion-exchange process[1] so that shorter range effects are more important than in the $NN$ interaction. In addition, the presence of the $\Lambda N - \Sigma N$ coupling influences the nature of the interaction. The mass difference between $\Lambda$ and $\Sigma$ is about 80 MeV and this coupling can cause significant effects in hypernuclei. Unfortunately, these features of the $YN$ interaction have not been well examined, mainly because $YN$ scattering data are scarce. Even $S$-wave $\Lambda N$ scattering lengths have yet to be determined.

Since 1993, we have analyzed the hypertriton[2], the lowest-mass hypernucleus, using various meson-theoretical $YN$ interactions. These analyses have clarified properties of the $S$-wave $YN$ interactions as well as confirmed the structure of the hypertriton as an extremely loosely bound system. Our findings are twofold: First, the binding energy is sensitive to the strengths of the $^3S_1$ and especially the $^1S_0$ $YN$ forces. The analyses combined with the $\Lambda N$ elastic total cross section data restrict the balance between the two force com-ponents to a small range. Second, the $\Lambda N - \Sigma N$ coupling plays the decisive role in the binding of the hypertriton. The expectation value of the $\Lambda N - \Sigma N$ coupling potential amounts to about half of the total $YN$ potential.

Our another interest is in the behavior of the $\Lambda N - \Sigma N$ coupling around the $\Sigma N$ threshold. Predictions of the $\Lambda N$ elastic total cross sections based on various $YN$ interaction models show an enhancement just around the $\Sigma N$ threshold. Reference [3] demonstrates that this is not a simple threshold effect but is caused by a $t$-matrix pole around the threshold. This enhancement is being investigated by photo- and electroproduction processes of a $K$ meson and a hyperon on the deuteron and $^3$He targets. We are pursuing such studies [4], and report a part of them in this conference (See H.Yamamura). This paper briefly surveys our recent results.

## 2  Analyses of the hypertriton

We solve the exact Faddeev equations for the coupled $\Lambda NN - \Sigma NN$ system. The formulation has appeared in several articles[2], to which we would like to refer the reader. For the $YN$ system, various modern meson-theoretical interactions are used, the Jülich $\tilde{A}$[5] model, and the soft core models of the Nijmegen group, NSC89[6] and NSC97[7]. The model NSC97 has six different versions, namely a, b, c, d, e and f, which contain varying relative strengths between the $^1S_0$ and $^3S_1$ force components. Among these models, only NSC89 and NSC97f bind the hypertriton at the correct binding energy. (The experimental $\Lambda$ separation energy is $0.13 \pm 0.05$ MeV.) The Jülich $\tilde{A}$ interaction cannot generate a bound state. This, as shown below, originates from a defect in their $^1S_0$ force. For the $NN$ sector, we use the Paris, Bonn B and Nijmegen93 interactions, but the results above do not depend on the choice of these potentials. It is the $^1S_0$ and $^3S_1 - ^3D_1$ partial waves in the $NN$ and $YN$ subsystems that dominate the hypertriton binding energy. The convergence with increasing number of partial waves is demonstrated in Ref. [2].

Figure 1(a) illustrates baryon-baryon correlation functions in the hypertriton, the probability to find the two particles at a distance $r$. The $NN$ correlation function is quite similar to the one for the deuteron which demonstrates that the $NN$ state in the hypertriton is close to the deuteron. The correlation functions multiplied by $r^2$ are displayed in Fig. 1(b), showing that the $\Lambda N$ correlation function has a large range. (The functions $\rho$ are normalized as $\int r^2 \rho_{NN} = 1$ and $\int r^2 [\rho_{N\Lambda} + \rho_{N\Sigma}] = 1$.) Thus, as expected with the small $\Lambda$ separation energy, the hypertriton is a loosely bound system where the $\Lambda$ barely clings to a deuteron.

Table 1 shows expectation values for the kinetic and potential energies in the hypertriton. Using the Njimegen93 $NN$ and NSC89 $YN$ potentials we see that the $NN$ subsystem is less bound than in the deuteron, and the relative kinetic energy of the $\Lambda$ with respect to the two nucleons $<T_{\Lambda-NN}>$ is larger than its potential energy $<V_{\Lambda N,\Lambda N}>$. It is due to the $\Lambda - \Sigma$ conversion that the hyperon is bound to the two nucleons. The expectation value $<V_{\Lambda N,\Sigma N}>$ amounts to more than half of the total $YN$ potential energy. In Table 2, we show the contributions from the $^1S_0$ and $^3S_1-^3D_1$ components to various $YN$ potential energies. The $^1S_0$ component dominates the expectation value of the direct potential $<V_{\Lambda N,\Lambda N}>$, while the $^3S_1-^3D_1$ component dominates the coupling potential $<V_{\Lambda N,\Sigma N}>$.

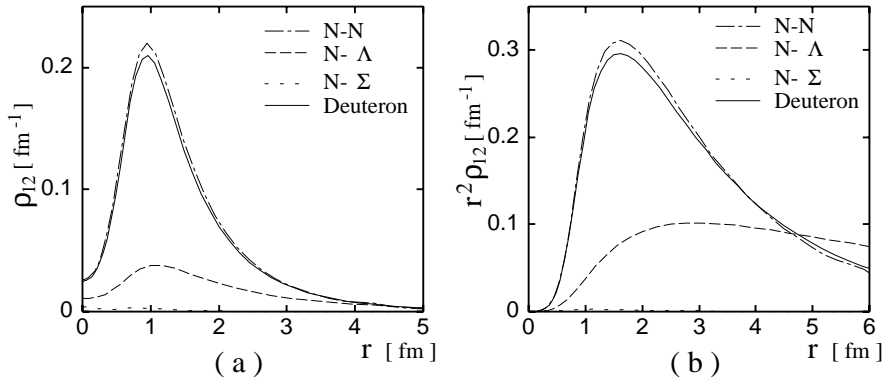

**Figure 1.** (a) Correlation functions for the $NN$, $\Lambda N$, and $\Sigma N$ pairs in the hypertriton. The correlation function in the deuteron is also shown for comparison. (b) Correlation functions multiplied by $r^2$.

**Table 1.** Expectation values for the various kinetic and potential energies in the hypertriton using the NSC89 $YN$ and Nijmegen93 $NN$ interactions.

| $<T_{NN}>$ | $<T_{\Lambda-NN}>$ | $<T_{\Sigma-NN}>$ | |
|---|---|---|---|
| 20.57 | 2.28 | 0.83 | |

| $<V_{NN}>$ | $<V_{\Lambda N,\Lambda N}>$ | $2<V_{\Lambda N,\Sigma N}>$ | $<V_{\Sigma N,\Sigma N}>$ |
|---|---|---|---|
| $-22.32$ | $-1.65$ | $-2.05$ | $-0.02$ |

**Table 2.** Contributions from the $^1S_0$ and $^3S_1-^3D_1$ force components to the expectation values of the $YN$ potential energies

| | $<V_{\Lambda N,\Lambda N}>$ | $2<V_{\Lambda N,\Sigma N}>$ | $<V_{\Sigma N,\Sigma N}>$ |
|---|---|---|---|
| $^1S_0$ | $-1.67$ | $-0.40$ | 0.03 |
| $^3S_1-^3D_1$ | 0.02 | $-1.61$ | $-0.06$ |

Let us discuss the strengths of the $^1S_0$ and $^3S_1$ force components in relation to the restriction imposed by the binding of the hypertriton. In Fig. 2, the $\Lambda N$ elastic total cross sections are shown for the various force models. All of the curves are adjusted to the sparse experimental data which have large error bars.

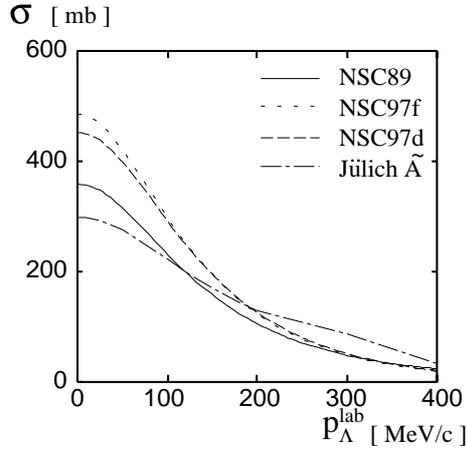

**Figure 2.** The $\Lambda N$ elastic total cross sections $\sigma$ predicted by the various $YN$ interactions as a function of the $\Lambda$ lab momentum.

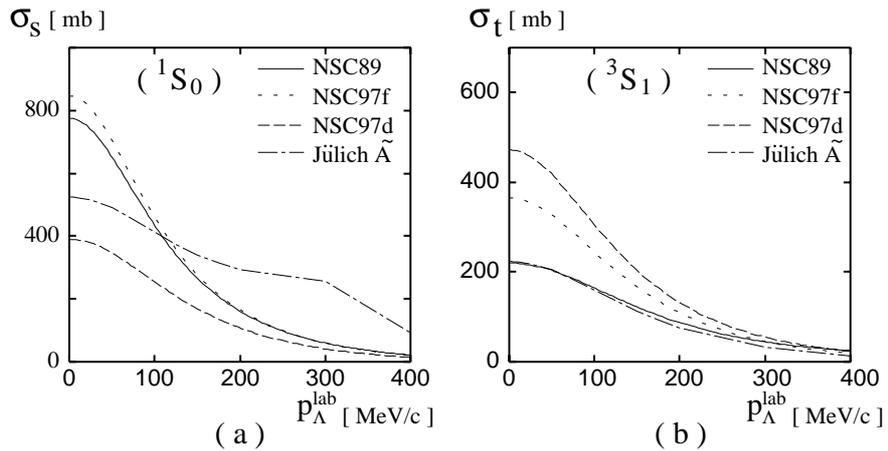

**Figure 3.** Partial $\Lambda N$ elastic total cross sections (a) for $^1S_0$, and (b) for $^3S_1$, which are defined as $\sigma = \frac{1}{4}\sigma_s + \frac{3}{4}\sigma_t$.

Among these potentials, NSC89 and NSC97f bind the hypertriton. The NSC97 potential, as mentioned, has six versions with different magnitudes between the $^1S_0$ and $^3S_1$ force components. Here, we take NSC97d as an example which does not bind the hypertriton. In Figs. 3(a) and 3(b), the separate contributions from the $^1S_0$ and $^3S_1$ states to the $\Lambda N$ elastic total cross sections are shown. They are defined as $\sigma = \frac{1}{4}\sigma_s + \frac{3}{4}\sigma_t$. As is seen in Figs. 3(a) and 3(b), the two potentials with the largest $^1S_0$ and small $^3S_1$ cross sections at low energies reproduce the hypertriton. From that we deduce that the scattering lengths are restricted to within $-2.7 \sim -2.4$ fm for $^1S_0$ and $-1.6 \sim -1.3$ fm for $^3S_1$.

## 3  Electromagnetic probes for the $YN$ and $YNN$ systems

$YN$ scattering experiments would be the best way to clarify the properties of the $YN$ interaction, but it is difficult to carry them out experimentally under present circumstances. Alternative and hopefully equally promising ways are experiments with electromagnetic probes such as $\gamma + d \to K^+ + Y + N$ and $e + d \to e' + K^+ + Y + N$, and similar experiments on $^3$He. One can obtain the information on the $YN$ interaction by analyzing the final state interactions among $YN$ or $YNN$. An inclusive $d(e, e'K^+)YN$ experiment has already been performed at TJNAF at Hall C, while the data for $d(\gamma, K^+Y)N$ are being analyzed in Hall B.

What dominates the $\Lambda N$ elastic total cross section curve is an enhancement around the $\Sigma N$ threshold which is predicted by all force models. We have investigated the amplitude for the $\Lambda N - \Sigma N$ system and found that it has a pole close to that threshold in the complex momentum plane. This pole will have a significant influence on various observables in the reactions mentioned above.

Beginning with photoproduction, we recently studied the inclusive $d(\gamma, K^+)YN$ and exclusive $d(\gamma, K^+Y)N$ processes for $\theta_K = 0°$ and found sizeable effects of the $YN$ final state interaction. This is reported by H.Yamamura in this conference. Even more pronounced are the final YN interaction effects on double polarization observables involving the recoil hyperon polarization along with a circularly polarized incoming photon. We plan to study the $e + d \to e' + K^+ + Y + N$ and $\gamma + ^3$He$\to K^+ + Y + N + N$ processes next. In the latter case, the $^3_\Lambda$H bound state just below the $\Lambda d$ threshold (0.13 MeV) is expected to yield significant effects on observables near threshold.

## 4  Summary

We have analyzed the hypertriton using various meson-theoretical $YN$ interactions. Among them, only NSC89 and NSC97f give the correct binding energy. The hypertriton is dominated by the $^1S_0$ and $^3S_1 - ^3D_1$ partial waves in the $NN$ and $YN$ subsystems, and has a structure such that a $\Lambda$ clings to a deuteron as expected from the small $\Lambda$ separation energy. The $\Lambda N - \Sigma N$ coupling plays the decisive role in this binding. The expectation value $<V_{\Lambda N, \Sigma N}>$ amounts to about half of the total potential energy $<V_{YN}>$. The strengths of the $^1S_0$ and $^3S_1$ $YN$ forces can be restricted by the combined analyses of the hypertriton and the $\Lambda N$ elastic cross section data. The scattering lengths have to lie within the intervals $-2.7 \sim -2.4$ fm for $^1S_0$ and $-1.6 \sim -1.3$ fm for $^3S_1$.

For further study of the $YN$ interaction, we believe experiments with electromagnetic probes are promising. We have studied the $\gamma + d \to K^+ + Y + N$ processes and found sizeable effects of the $YN$ final state interaction. The anal-yses of $e + d \rightarrow e' + K^+ + Y + N$ and $\gamma + {}^3\text{He} \rightarrow K^+ + Y + N + N$ processes are underway.